\documentclass[preprint,12pt]{elsarticle}

\usepackage{amssymb}

\journal{Nuclear Instruments and Methods in Physics Research A}

\begin{document}

\begin{frontmatter}

\title{Reconstruction efficiency and discovery potential of a Mediterranean neutrino telescope: A simulation study using the Hellenic Open University Reconstruction \& Simulation (HOURS) package}

\author{A. G. Tsirigotis\corref{cor1}}
\cortext[cor1]{Corresponding author. Tel: +302610367517; Fax: +302610367528}
\ead{tsirigotis@eap.gr}
\author{A. Leisos}
\author{S. E. Tzamarias}
\address{Physics Laboratory, School of Science \& Technology, Hellenic Open University}
\author{On behalf of the KM3NeT Consortium}

\begin{abstract}
We report on the evaluation of the performance of a Mediterranean very large volume neutrino
telescope. We present results of our studies concerning the capability of the telescope in
detecting/discovering galactic (steady point sources) and extragalactic, transient (Gamma Ray Bursts)
high energy neutrino sources as well as measuring ultra high energy diffuse neutrino fluxes. The
neutrino effective area and angular resolution are presented as a function of the neutrino energy, and the
background event rate (atmospheric neutrinos and muons) is estimated. The discovery potential of the
neutrino telescope is evaluated and the experimental time required for a significant discovery of
potential neutrino emitters (known from their gamma ray emission, assumedly produced by hadronic
interactions) is estimated. For the simulation we use the HOU Reconstruction \& Simulation (HOURS)
software package.
\end{abstract}

\begin{keyword}
  Neutrino Telescope \sep Event Simulation \sep Detector Evaluation
  \PACS 95.55.Vj
\end{keyword}

\end{frontmatter}

%%%%%%%%%%%%%%%%%%%%%%%%%%%%%%%%%%%%%%%%%%%%%%%%%%%%%%%%%%%%%%%%%%%%%%%%%%%%%%%%%%%%%%%%%%%%%%%%%%%%%%%%
\section{Introduction}
\label{intro}
KM3NeT (km$^3$ Neutrino Telescope) is planned to be a deep-sea multidisciplinary observatory in the Mediterranean Sea that will provide innovative science
opportunities spanning Astroparticle Physics and Earth and Sea Science \cite{km3netcdrtdr}. 
Various astrophysical sources are expected to produce high-energy neutrinos that
may be detected with KM3NeT.
The observation of even a handfull of events emanating from a cosmic source could provide 
usefull information.
The existence of these neutrino sources will be proved and more importantly knowledge 
of their behavior, which cannot be acquired by other means, will be gained.

In this work we study the performance of KM3NeT, by estimating the sensitivity in detecting cosmic neutrino fluxes from astrophysical neutrino
point or extended sources.
In Sections \ref{detector} and \ref{simulation} we describe the neutrino telescope configuration under study and the simulation framework.
The telescope's performance in detecting neutrinos from astrophysical sources is described in Section \ref{performance}, 
while the experimental time required for a significant discovery of potential Galactic neutrino emitters is estimated in Section \ref{galactic}.
%%%%%%%%%%%%%%%%%%%%%%%%%%%%%%%%%%%%%%%%%%%%%%%%%%%%%%%%%%%%%%%%%%%%%%%%%%%%%%%%%%%%%%%%%%%%%%%%%%%%%%%%
\section{Detector description}
\label{detector}
KM3NeT will consist of several hundreds of vertical structures (Detection Units - DUs), which carry photo-sensors
and devices for calibration and environmental measurements, arranged vertically on
Storeys. Each Storey will support one or two photo-sensors. The photo-sensor unit
is a digital optical module (DOM) consisting of a 17-inch diameter pressure resistant
glass sphere housing 31 3-inch photomultiplier (PMT) tubes, their high-voltage bases
and their interfaces to the data acquisition system with nanosecond timing precision \cite{km3netppm}.
The segmentation of the photocathode area in such a Multi-PMT Optical Module 
will aid in distinguishing single-photon from multi-photon hits, and thus provide a better 
optical background rejection and trigger efficiency.
The front-end electronics is based on the use of the time over threshold (ToT) as the main signal processing technique
\cite{km3netppm,bourlis}.
The ToT technique is based on the use of a time to digital converter (TDC) that performs time-tagging of the leading and
trailing edge of the PMT signal above a certain voltage threshold.
The signal goes through a comparator that compares
it against the desired threshold. The output of the comparator is
then fed to the TDC that performs the time-tagging of the leading
and trailing edge. These values are subsequently used for the
reconstruction of the pulse shape and its charge.

In the present study the telescope layout considered is the one optimized during the KM3NeT Design Study, 
exhibiting optimal sensitivity in discovering astrophysical point sources emitting neutrinos with an
energy spectrum of $E^{-2}$ and a high energy cut-off or without one. 
According to this layout, the KM3NeT detector will consist of 12320 DOMs distributed over 308 DUs. 
Each DU consists of 20 Storeys, and each Storey consists of a bar with one DOM at
either end. The distance between the centers of the DOMs is 6~m. The distance between Storeys
is 40~m, while the position of the lowest Storey is 100~m above the seabed.
The bars have a direction orthogonal to their neighbors.
The distribution of the positions of the detection units on the seabed (the so-called
footprint) is homogeneous. The footprint forms a roughly circular shape and has a typical DU
density corresponding to an average distance between neighboring detection units of about 180~m.
The total instrumented volume of the detector is 5.8~km$^3$.
In the present study the depth of the seabed is considered to be 3500~m.
%%%%%%%%%%%%%%%%%%%%%%%%%%%%%%%%%%%%%%%%%%%%%%%%%%%%%%%%%%%%%%%%%%%%%%%%%%%%%%%%%%%%%%%%%%%%%%%%%%%%%%%%
\section{Simulation Framework}
\label{simulation}
For the evaluation of the KM3NeT performance in detecting high energy astrophysical neutrinos the HOURS 
(Hellenic Open University Reconstruction \& Simulation) physics analysis package was used. 
HOURS comprises a realistic simulation package of the detector response, including an accurate description 
of all the relevant physical processes, the production of signal and background as well as several 
analysis strategies for triggering and pattern recognition, event reconstruction, tracking and energy estimation \cite{HOURS}.
In HOURS the Kalman filter is used as a recursive track fitting method, as described in \cite{kalman}.
Using the Kalman filter technique a large number of candidate tracks are found for each event.
The best candidate (hereafter reconstructed track) is chosen using the likelihood value corresponding to the directional sensitivity of the Multi-PMT DOM
\cite{kalman} and the probability density function of the arrival time residuals of the hits\footnote{In the calculation
of the likelihood value the hits are common for all candidate tracks.}.
The available quality cuts for the final event selection are the number of candidate tracks found, 
the likelihood value and number of hits of the reconstructed track, and the number of candidate tracks 
with an angular deviation to the reconstructed track less than $5^\circ$.

For the present study we have simulated the response of the detector configuration described in Section \ref{detector} 
to a generic neutrino flux distributed isotropically (in a $4\pi$ solid angle) on the Earth's atmosphere. 
We assumed that the energy distribution of these neutrinos follows a power law spectrum, 
with a spectral index of $-2.0$, in the range of $15~\mathrm{GeV}-100~\mathrm{PeV}$
\footnote{The lower energy limit in the event generation has been chosen to be of the order of the neutrino energy 
threshold of the simulated detector which is about 30~GeV (see also Fig.~\ref{km3neteffarea}).
In the event generation the cross section of the neutrino interaction is taken into account.
The energy spectrum of the interacting neutrinos that produce a detectable signal has a peak at 16~TeV,
while 90\% of them are in the range of $100~\mathrm{GeV}-1~\mathrm{PeV}$.}.

Using the expected neutrino flux for each point or extended source, 
we estimated the number of neutrino signal events per observation year.
The estimation was based on reweighting each of the signal events 
(produced by a cosmic neutrino) according to the energy of the 
parent neutrino and taking into account the functional form of the generic flux, 
used in the Monte Carlo event production, and the energy dependence of the source spectrum.

In our studies we have included the contribution of two main background sources, 
namely atmospheric neutrinos and energetic muons. In order to simulate the first background source we used the Bartol flux parametrization \cite{bartol}, 
including the high energy component from charm interactions (prompt neutrinos). For the atmospheric muon component we have used the 
MUPAGE package \cite{mupage} to simulate the generated muons from energetic Extensive Air Showers.
Although we have generated a large statistical sample of cosmic and atmospheric neutrinos, the atmospheric muon background corresponds to only 1 h equivalent of integrated flux (lifetime), due to CPU time constrains.

The search for cosmic neutrino signals is based on statistical techniques that are appropriate for the detection of small numbers of
events. The discovery potential of the neutrino telescope is evaluated using the model discovery potential (MDP) calculated for
a level of significance $5\sigma$ and probability to make a discovery 50\%. In this case the quality cuts to the reconstructed tracks 
are optimized in order to obtain the least signal necessary to claim a discovery.
When no significant signal is observable, limits are set using the model rejection factor (MRF). 
In this case, the quality cuts to the reconstructed tracks are optimized in order
to obtain the lowest possible expected upper limit for the experiment, assuming that no true signal is present. 

For point-like (or with a small angular size) source searches the sensitivity of the detector has
been obtained applying a ``binned'' method where the sky is divided in bins of declination and right ascension and the 
numbers of events detected per bin are analyzed. The parameters that are optimized in order to minimize the MRF and MDP are the size of
the search cone around the source and the quality cuts to the reconstructed tracks. 
The average number of signal events from the source, at a given declination, and the average number of background events
inside the selected search cone centered on the source direction is estimated.
%%%%%%%%%%%%%%%%%%%%%%%%%%%%%%%%%%%%%%%%%%%%%%%%%%%%%%%%%%%%%%%%%%%%%%%%%%%%%%%%%%%%%%%%%%%%%%%%%%%%%%%%
\section{Telescope performance}
\label{performance}
The observation of point-like sources of neutrinos would bring unique new insights
on the nature of cosmic accelerators and resolve the enigma of the origin of cosmic
rays. Observations by gamma ray telescopes have revealed many astrophysical objects, 
in which high-energy processes at and beyond the TeV level take place. 
However, measurements with gamma rays alone cannot clearly distinguish whether the 
accelerated particles are leptons or hadrons. 
Only the observation of neutrinos from a source can unambiguously establish the 
hadronic nature of that source.
%%%%%%%%%%%%%%%%%%%%%%%%%%%%%%
\begin{figure}
\begin{center}
\includegraphics[width=0.8\textwidth]{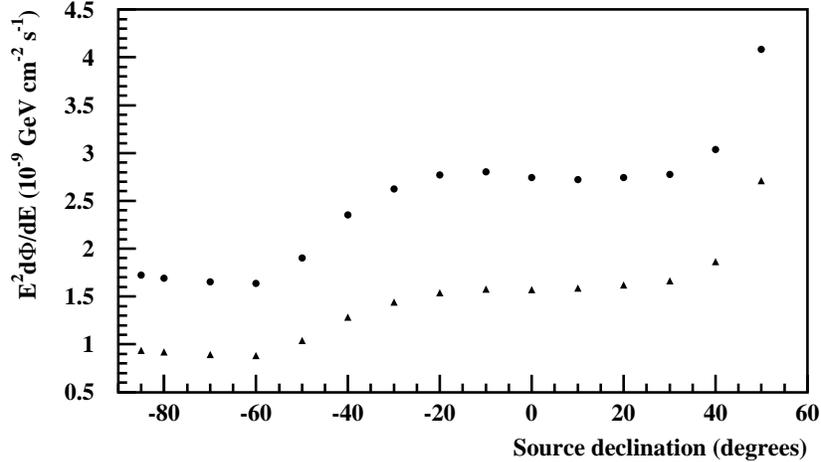}
\end{center}
\caption{The KM3NeT flux sensitivity (triangles) and discovery flux (circles)
for one year of observation and assuming an $E^{-2}$ source spectrum without cutoff.}
\label{km3netsens}
\end{figure}
%%%%%%%%%%%%%%%%%%%%%%%%%%%%%

The sensitivity of the detector to neutrino point sources, based on one year of data taking, 
is shown in Fig.~\ref{km3netsens} as a function of the declination. 
The detector performance is presented as the flux that can be excluded at 90\% CL 
(flux sensitivity) and the flux that can be detected at $5\sigma$ with 50\% probability 
(discovery flux). This calculation assumes a neutrino energy spectrum proportional to $E^{-2}$ 
with no energy cutoff.
%%%%%%%%%%%%%%%%%%%%%%%%%%%%%%%
\begin{figure}
\begin{center}
\includegraphics[width=0.8\textwidth]{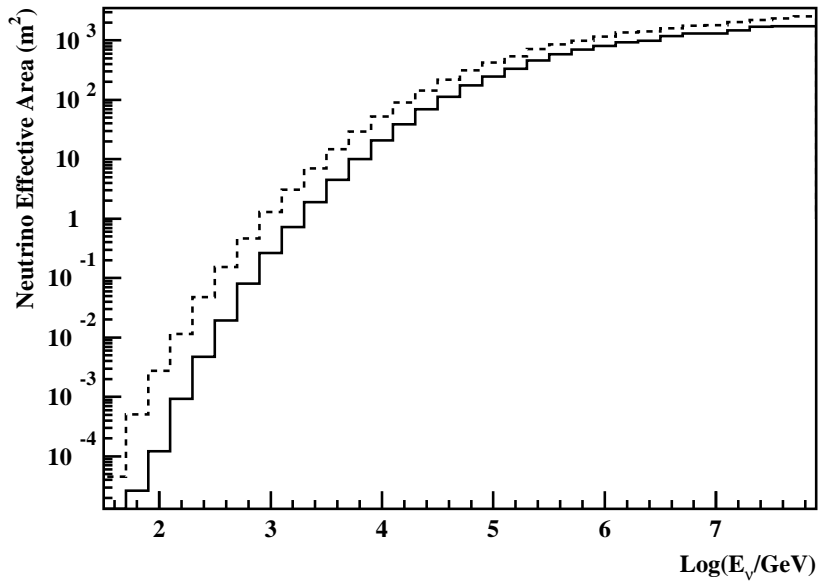}
\end{center}
\caption{The effective area of the neutrino telescope as a function of the neutrino
energy, for the triggered (dashed line) and the reconstructed (solid line) events. 
The quality cuts applied to the reconstructed tracks are the same
as those applied in evaluating the detector's flux sensitivity seen in Fig.~\ref{km3netsens}.}
\label{km3neteffarea}
\end{figure}
%%%%%%%%%%%%%%%%%%%%%%%%%%%%%%%

In Fig.~\ref{km3neteffarea} the Neutrino effective area as a function of the neutrino energy is shown for the triggered and reconstructed
events.
The event trigger used in the present study is minimal and is based on five (or more) L1 hits on different DOMs due to the signal from a neutrino interaction 
or atmospheric muons. An L1 hit is defined as a local 
coincidence, within a time window of 10~ns, of two (or more) photons detected within the same optical module by different PMTs. 
%%%%%%%%%%%%%%%%%%%%%%%%%%%%%
\begin{figure}
\begin{center}
\includegraphics[width=0.8\textwidth]{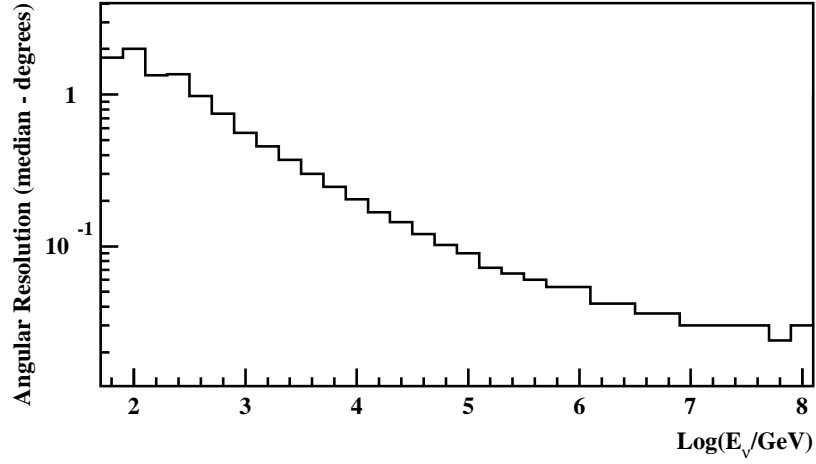}
\end{center}
\caption{The angular resolution as a function of the neutrino energy. 
The quality cuts applied to the reconstructed tracks are the same
as those applied in evaluating the detector's flux sensitivity seen in Fig.~\ref{km3netsens}.}
\label{km3netangl}
\end{figure}
%%%%%%%%%%%%%%%%%%%%%%%%%%%%%%%
The angular resolution, in reconstructing the neutrino direction, is shown in Fig.~\ref{km3netangl} as a function of the neutrino energy.
%%%%%%%%%%%%%%%%%%%%%%%%%%%%%%%
\begin{figure}
\begin{center}
\includegraphics[width=0.8\textwidth]{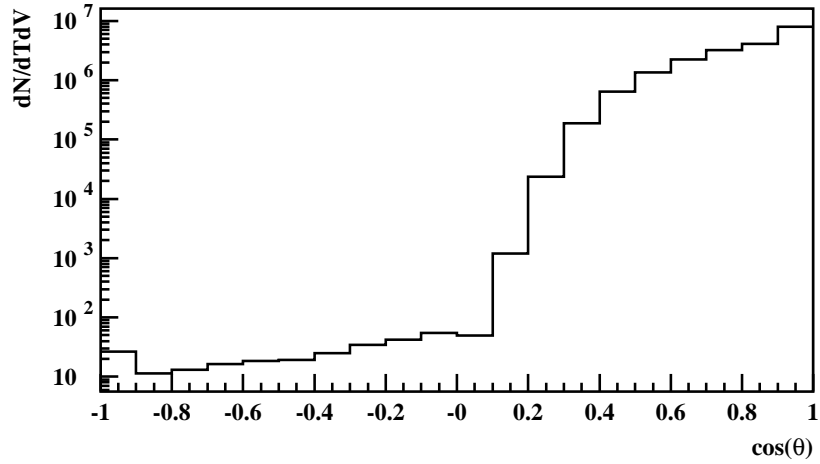}
\end{center}
\caption{The number of background events per unit time (days) and per unit solid angle as a function of the zenith angle.}
\label{km3netevrate}
\end{figure}
%%%%%%%%%%%%%%%%%%%%%%%%%%%%%%%%%%

The rate of background events due to atmospheric neutrinos and atmospheric muons, 
remaining after the application of quality cuts to the reconstructed tracks, is presented in Fig.~\ref{km3netevrate}.
The peak in the vertically up-going and down-going ($cos(\theta)=-1$ and $cos(\theta)=1$) bins is due to the
increased efficiency of the detector for low energy vertical muon tracks passing close to one Detection Unit.
The number of reconstructed upcoming atmospheric neutrinos per day is about 170, 
while the number of reconstructed downcoming atmospheric muon events is about 20 million per day.

Gamma Ray Bursts are also potential very high energy neutrino emitters according to
the fireball model \cite{grbs}. High energy neutrinos from prompt emission consistent with the
detected gamma rays are expected to arrive within a short time window ($2 - 1000$~s) \cite{grbs2}. The narrow
time window results in reduced background noise and with the combination of an
appropriate cut on the reconstructed energy of the neutrino induced muon, the detection of down-going
GRB neutrinos is feasible \cite{tools}.
Studies have shown that KM3NeT could observe about 5 neutrinos per year
from GRBs (according to the optimistic model of \cite{grbs}, and assuming that 300 GRBs per year are observed by space born
experiments).

The ultra high energy neutrinos from: a) a multitude of objects such as Active
Galactic Nuclei or GRBs, and b) from the interaction of cosmic rays with intergalactic
matter and radiation or even with the cosmic microwave background, are expected to
form an isotropic diffuse flux. Without the possibility of using a tight angular cut for
reducing the background of atmospheric neutrinos, diffuse neutrino flux searches have
to rely on a cut on the reconstructed muon energy, $E_{\mu reco}$.
The method, used in HOURS, to estimate the energy of a reconstructed muon, is described in \cite{HOURS}. 
For this telescope configuration, the log-energy resolution, $\Delta (log E)$, is described by the 
empirical formula, $\Delta (log E) = 0.32 + 0.082 \cdot arctan (10 - 2.42 \cdot log E)$,
where $E$ is the true muon energy in GeV. The log-energy resolution varies from 0.42 at 1~TeV to 0.23
at 100~TeV reaching the value 0.2 at 1~PeV.
The sensitivity of KM3NeT, for one year of observation time, 
to an isotropic diffuse neutrino flux with a spectrum proportional to $E^{-2}$, 
and an energy cut $E_{\mu reco} >500$~TeV has 
been estimated to be $3\times10^{-9}$~(GeV$^{-1}$~cm$^{-2}$~s$^{-1}$~sr$^{-1}$).
%%%%%%%%%%%%%%%%%%%%%%%%%%%%%%%%%%%%%%%%%%%%%%%%%%%%%%%%%%%%%%%%%%%%%%%%%%%%%%%%%%%%%%%%%%%%%%%%%%%%%%%%
\section{Observation of Galactic neutrino sources}
\label{galactic}
Supernova remnants (SNR) of the shell type are the most probable sources of cosmic rays in the Galaxy. 
The material ejected during the explosion forms shock waves when it propagates into the interstellar matter. 
Particles are assumed to be accelerated in these shock waves, which can persist for several thousand years. 
The shell-type SNRs with the most intense gamma rays fluxes are RX J0852.0-4622 (Vela Junior) and 
RX J1713.7-3946. These sources have an angular size larger than the resolution of the neutrino telescope.
Moreover, they are generally expected to have a cut-off in their energy spectra in the range $1-10~\mathrm{TeV}$.
However, the neutrino telescope layout, optimized during the KM3NeT Design Study, exhibits
optimal sensitivity in discovering astrophysical sources emitting neutrinos with an
energy spectrum of $E^{-2}$ and a high (or without any) energy cut-off \cite{optimize}. Such a detector
has lower sensitivity in detecting galactic neutrinos in the energy range of $1-10~\mathrm{TeV}$.

Assuming that the dominating mechanism of gamma ray production is hadronic, 
the most luminous gamma ray Galactic source, the SNR RX J1713.7-3946, is estimated to emit neutrinos with a flux \cite{rxjflux}
\begin{eqnarray}
\Phi(E)=16.8 \times 10^{-15} \left[\frac{E}{\mathrm{TeV}}\right]^{-1.72} \times \nonumber \\
e^{-\sqrt{\frac{E}{2.1~\mathrm{TeV}}}}~\mathrm{GeV^{-1}~s^{-1}~cm^{-2}} .
\end{eqnarray}
Such a source can be detected in 13.1 years of experimental running time of the previously described detector, 
as is shown in Table \ref{d5srxj}.
By reducing the average distance between the DUs from 180~m to 130~m the number of years required 
for a $5\sigma$ discovery with a 50\% probability is reduced to 9 years.
Furthermore, if we replace each DU by two strings of DOMs, and redistribute them homogeneously with an average 
distance between them of 100~m, the detector running time required for the discovery of this Galactic source can be reduced to about 7 years. 
%%%%%%%%%%%%%%%%%%%table of detectors fom
\begin{table*}[ht]
  \centering
  \caption{Number of years required for a significant discovery of the SNR RX J1713.7-3946 for various KM3NeT layouts.}
          {%\footnotesize
            \hfill{}
            \begin{tabular}{ c c c c c }
              \hline
              Number        & Average distance     & Number of    & Instrumented    &  Years for  \\ 
              of DUs        & between DUs (m)      & DOMs per DU  & volume (km$^3$) &  discovery  \\ \hline
              308           & 180                  & 40           &   5.8           &  13.1                        \\ 
              308           & 130                  & 40           &   3.0           &  9.05                        \\ 
              616           & 180                  & 20           &   11.6          &  13.3                        \\ 
              616           & 130                  & 20           &   6.1           &  7.61                        \\ 
              616           & 100                  & 20           &   3.6           &  7.15                        \\
              \hline
          \end{tabular}}
          \hfill{}
          \label{d5srxj}
\end{table*}
%%%%%%%%%%%%%%%%%%%%%%%%%%%%%%%%%%%%%%%%%%%

Preliminary studies have shown \cite{underprep}
that the most luminous Galactic sources can be detected in less than 5 years 
(e.g. using the telescope configuration of 616 strings of DOMs described above)
of detector running time by applying more sophisticated experimental and data analysis techniques, 
that take into account:
(a) the track reconstruction resolution and the reconstructed energy of the neutrino induced muon 
on a track by track basis, as described in \cite{shape},
(b) the known source direction as described in \cite{dir}, and 
(c) the source morphology \cite{underprep}.
\section{Conclusions}
In this work we have studied, using the HOURS package, the
performance of the Mediterranean very large volume neutrino
telescope, KM3NeT, in discovering Galactic and extragalactic neutrino sources.
KM3NeT will cover most of the sky with unprecedented sensitivity and in the first few years
of operation will unambiguously discover neutrinos from many promising Galactic candidate sources.
%%%%%%%%%%%%%%%%%%%%%%%%%%%%%%%%%%%%%%%%%%%%%%%%%%%%%%%%%%%%%%%%%%%%%%%%%%%%%%%%%%%%%%%%%%%%%%%%%%%%%%%%
\section*{Acknowledgments}
The KM3NeT project is supported by the EU in FP6 under Contract no. 011937
and in FP7 under Grant no. 212525
%%%%%%%%%%%%%%%%%%%%%%%%%%%%%%%%%%%%%%%%%%%%%%%%%%%%%%%%%%%%%%%%%%%%%%%%%%%%%%%%%%%%%%%%%%%%%%%%%%%%%%%%
\bibliographystyle{elsarticle-num}

\end{document}